\documentclass[useAMS,usenatbib]{mn2e}

\usepackage{graphicx}

\title[Inverse Hubble Flows]{Inverse Hubble Flows in Molecular Clouds}
\author[Toal\'{a} et al.]{Jes\'{u}s A. Toal\'{a}$^{1}$\thanks{E-mail:toala@iaa.es}, 
Enrique V\'{a}zquez-Semadeni$^{2}$, Pedro Col\'{i}n$^{2}$ and Gilberto C. G\'{o}mez$^{2}$\\
$^{1}$Instituto de Astrof\'{i}sica de Andaluc\'{i}a, IAA-CSIC, Glorieta de la 
Astronom\'{i}a s/n, 18008 Granada, Spain\\
$^{2}$Centro de Radioastronom\'{i}a y Astrof\'{i}sica, UNAM, Campus Morelia, Apartado Postal 3-72, 
58089, Morelia, Michoac\'{a}n, Mexico}

\def\gtsima{$\; \buildrel > \over \sim \;$}    % Use in text mode
\def\gtrsim{\lower.5ex\hbox{\gtsima}}           % Use in math mode
\def\lesssim{\lower.5ex\hbox{\ltsima}}           % Use in math mode
\def\ltsima{$\; \buildrel < \over \sim \;$}    % Use in text mode

\newcommand{\rr} {{\bf r}}
\newcommand{\taunl} {\tau_{\rm nl}}
\newcommand{\tauff} {\tau_{\rm ff}}

\newcommand{\aap} {A\& A}
\newcommand{\apj} {ApJ}
\newcommand{\apjl} {ApJL}
\newcommand{\deltac} {\delta_{\rm C}}

\newcommand{\deltamc} {\delta_{\rm MC}}
\newcommand{\etal} {et al.}
\newcommand{\mnras} {MNRAS}
\newcommand{\VS} {V\'azquez-Semadeni}

\begin{document}

%\date{Accepted 1988 December 15. Received 1988 December 14; in original form 1988 October 11}

%\pagerange{\pageref{firstpage}--\pageref{lastpage}} \pubyear{2002}

\maketitle

\label{firstpage}

\begin{abstract}

  Motivated by recent numerical simulations of molecular cloud (MC)
  evolution, in which the clouds engage in global gravitational
  contraction, and local collapse events culminate significantly
  earlier than the global collapse, we investigate the growth of
  density perturbations embedded in a collapsing background, to which
  we refer as an {\it Inverse Hubble Flow} (IHF).  We use the standard
  procedure for the growth of perturbations in a universe that first
  expands (the usual Hubble Flow) and then recollapses (the IHF).  We
  find that linear density perturbations immersed in an IHF grow
  faster than perturbations evolving in a static background (the
  standard Jeans analysis).  A fundamental distinction between the two
  regimes is that, in the Jeans case, the time $\tau_\mathrm{nl}$ for a density
  fluctuation to become nonlinear increases without limit as its
  initial value approaches zero, while in the IHF case $\tau_\mathrm{nl} \le
  \tau_\mathrm{ff}$ always, where $\tau_\mathrm{ff}$ is the free-fall time of the
  background density.  We suggest that this effect, although moderate,
  implies that small-scale density fluctuations embedded in
  globally-collapsing clouds must collapse earlier than their parent
  cloud, regardless of whether the initial amplitude of the
  fluctuations is moderate or strongly nonlinear, thus allowing the
  classical mechanism of Hoyle fragmentation to operate in
  multi-Jeans-mass MCs.  More fundamentally, our results show that,
  contrary to the standard paradigm that fluctuations of all scales
  grow at the same rate in the linear regime, the hierarchical nesting
  of the fluctuations of different scales does affect their growth
  even in the linear stage.

\end{abstract}


\section{Introduction} \label{sec:intro}

The fragmentation of a local overdensity (a ``cloud'') in
a continuum is one of the fundamental problems in astrophysics, as it
underlies the formation of galaxy and star clusters. Over sixty years
ago, \citet{Hoyle1953} proposed a model in which stars form in a
fragmentation process {\it during} the collapse of a nearly isothermal
spherical cloud, based on the fact that the Jeans mass in an isothermal
medium \citep[or, more generally, in any polytropic medium with
polytropic exponent $\gamma < 4/3$; e.g.,] [] {Chandra61} decreases as
the density increases. Hoyle's mechanism was subsequently laid on firmer
mathematical grounds by \citet{Hunter1962, Hunter1964}. This point of
view prevailed until it was realized by
\citet{Tohline80} that the small-scale density fluctuations within a
cloud whose initial mass is close to the Jeans mass could not
grow faster than the cloud itself, because when thermal pressure is
non-negligible, it acts to slow down the collapse of smaller-scales
compared to the larger scales. Since then, it has been generally
accepted that molecular clouds (MCs) do not fragment by a Hoyle-like
mechanism, and other mechanisms, such as turbulent fragmentation
\citep[e.g.,] [] {MK04} might be at work.

However, recent numerical studies suggest that cold, dense clouds in the
interstellar medium (ISM) can form by intermediate- to large-scale
converging flows in the warm atomic medium, which are capable of
coherently triggering a phase transition to the cold atomic phase over
large regions in the gas \citep[e.g.,] [] {BP+99, HP99, WF00, HBB01,
KI02, AH05, Heitsch+05, Heitsch+06, VS+06}. This large-scale coherence
implies that the clouds can form already containing a large number of
Jeans masses \citep{VS+07}.  
%Moreover, the formation mechanism generates
%moderately supersonic turbulence in the condensed gas \citep{WF00, KI02,
%Heitsch+05, Heitsch+06, VS+06}, producing {\it nonlinear} density
%fluctuations in the cloud, which have shorter free-fall times than the
%whole cloud.
%Finally, the formation of clouds by coherent large-scale compressions
%implies that the morphology is not spherical, but rather flattened, and
%it has been recently shown that 
Therefore, the crucial assumption made by
\citet{Tohline80}, that the cloud's mass is near the Jeans mass,
is not necessarily fulfilled in collapsing MCs, thus making it relevant
to again consider the collapse of small-scale fluctuations within a
larger-scale object which is itself collapsing, for the fragmentation of
MCs.

The analysis of the collapse of density structures embedded within
larger ones that are also undergoing collapse may benefit from the tools
developed for studying the evolution of linear fluctuations in the
cosmological flow of dark matter. In this case, it is standard to
consider a pressureless, expanding Hubble flow, in which small-amplitude
density enhancements at a certain scale $L_0$ begin to retard their
expansion, until they eventually begin to collapse, at which point,
they are said to
``separate'' from the global expansion
\citep[e.g.,] [] {KT90}.  It is well known that,
in this case, the collapse of these regions proceeds more slowly (as a
power-law in time) than that of a fluctuation in a non-expanding medium
(which grows exponentially), because the global expansion counteracts
the global collapse. 
%However, if a full spectrum of density fluctuations
%is present in the flow, then necessarily some smaller-scale fluctuations
%(of scale, say, $L_1 < L_0$) will happen to be located inside the
%larger-scale collapsing structure, and those will be embedded in an {\it
%inverse Hubble flow} (IHF). 
Now, if being embedded in an expanding
(regular Hubble) flow reduces the growth rate of the large-scale
fluctuation, it is natural to expect that being embedded in a {\it
contracting} (inverse Hubble) flow should {\it enhance} the growth rate
of the fluctuations located inside the collapsing large-scale
structure. This could be thought of as an inescapable form of
nonlinearity, in the sense that the growth of one mode is linked to that
of another mode, since the growth of a small-scale fluctuation depends
on its being located within a larger-scale one, an effect that will be
active even when the small-scale fluctuation has a small (linear)
amplitude.

In this paper, we investigate this possibility, using a linear
perturbation analysis to study the growth of density fluctuations
located inside a larger-scale spherical fluctuation undergoing free-fall
collapse as well. Because, as dicussed above, the MC case may be adequately
described by means of a nearly pressureless regime, while the dark
matter is intrinsically so, we consider a pressureless regime here. The
solution for the density perturbation growth is then compared to that of
the classical Jeans' case for a static background, to show that the
initially linear density perturbations grow at a faster rate inside a
collapsing spherical cloud.
%because are being acreted by the collapsing parent cloud. 
%Thus, hierarchical fragmentation occurs
%not only in the case where we have nonlinear inhomogeneities but also 
%if the seeds are in the linear regime.

\section{The governing equations} \label{sec:gov_eqs}

\subsection{The contracting background} \label{sec:contr_bkgd}

%In general, the study of density instabilities can be applied to a
%cloud or to the cosmology case (expanding or collapsing universe). 

In what follows, we will consider the collapse of a {\it spherical}
cloud, embedded in a medium that is itself collapsing. Although MCs in
particular are known to strongly depart from a spherical symmetry
\citep[e.g.,] [] {Bally+87, Gutermuth+08, Myers09, Menshchikov+10,
Molinari+10}, and the collapse times for non-spherical clouds are known
to be longer than the standard free-fall time, which assumes this
geometry \citep{Toala+12, Pon+12}, the study of the spherical case will
allow us to compare with this standard timescale.

The collapse of a spherical cloud can be described using the standard
machinery applied for a contracting Universe, noting that the physical
differences lie in the definition of the Hubble parameter $H(t)$. In
both cases this can be written as $H(t) = \dot{a}(t)/a(t)$, where $a(t)$
is a suitable scale factor. For the cosmological case, $a$ is simply the
well-known scale factor, while for a spherical MC, $a$ can be
defined as $a(t)=R(t)/R_0$, where $R_{0}$ is the initial cloud's
radius. In both the MC and the cosmological cases, $a(t)$ satifies
Friedmann's equation (see Appendix A for the MC case)
\begin{equation}
\left(\frac{\dot{a}}{a}\right)^{2} + \frac{k}{a^{2}} = \frac{8}{3}\pi G
\rho. 
\label{eq:Friedmann}
\end{equation}

The case of a collapsing spherical cloud is mathematically equivalent to
the second half of the evolution of a closed ($k>0$), matter-dominated
universe that frst expands to a certain maximum scale factor and then
recollapses. So, in what follows, we will consider that the origin of
the time coordinate is the point of maximum expansion of such closed
Universe. Note that, during this contracting phase, $\dot a <0$.

\subsection{The linear analysis} \label{sec:lin}

The standard linear stability analysis for the growth of density
fluctuations in the case of an expanding (or contracting) universe
starts from the linearized equations of continuity and momentum
conservation, together with the Poisson equation, which read
\citep[e.g.,] []{KT90}
\begin{eqnarray}
\frac{\partial \rho_1}{\partial t}+\frac{3\dot a}{a}\rho_1+\frac{\dot
a}{a}({\bf r}  \cdot  {\bf \nabla})\rho_1+\rho_0{\bf \nabla} \cdot {\bf
v_1} &=& 0 \nonumber \\
\frac{\partial {\bf v_1}}{\partial t}+\frac{\dot a}{a}{\bf v_1}
+\frac{\dot a}{a}({\bf r}  \cdot {\bf \nabla}) {\bf v_1}
+\frac{v^2_s}{\rho_0}{\bf \nabla}\rho_1+{\bf \nabla}\varphi_1 &=& 0
\nonumber \\
\nabla^2\varphi_1 &=& 4\pi G\rho_1,
\end{eqnarray}
where $v_{s}$ denotes the sound speed, $\rr$ denotes the position
vector, and the variables (generically denoted $x$) have been
decomposed as $x = x_0 + x_1$, the subindex ``0'' denoting the
unperturbed quantities, and the subindex ``1'' denoting the
corresponding (small) fluctuation. The unperturbed quantities include
the background expansion (or contraction):
\begin{equation}
\rho_0 (t)=\rho_0(t_0)\, a^{-3}(t)~~~~~~{\bf v_0} =\frac{\dot R}{R}{\bf r}
~~~~~~{\bf \nabla}\varphi_0=\frac{4}{3}\pi G\rho_0 {\bf r},
\end{equation}
where $a(t)$ is the scale factor, obeying the Friedmann equation, eq.\
(\ref{eq:Friedmann}).

As is well known, for scales much larger than the Jeans scale, that is,
neglecting the pressure term, the equation for the perturbation growth
\citep[e.g.,] [] {Mukhanov05} is given by
\begin{equation}
\ddot{\delta} + 2H(t)\dot{\delta} - 4 \pi G \rho_{0} \delta =0,
\label{eq:deltas}
\end{equation}
where $\delta \equiv \rho_1/\rho_0 \equiv (\rho - \rho_0)/\rho_0$ is the
relative density fluctuation. A general solution to eq.\
(\ref{eq:deltas}) can be written as \citep{Mukhanov05}:
\begin{equation}
\delta = C_{1} H(t)\int \frac{dt}{a(t)^{2}H(t)^{2}} + C_{2}H(t),
\end{equation}
where $C_{1}$ and $C_{2}$ are constants.

\begin{figure}%[p]
\begin{center}
 \includegraphics[width=1.0\linewidth]{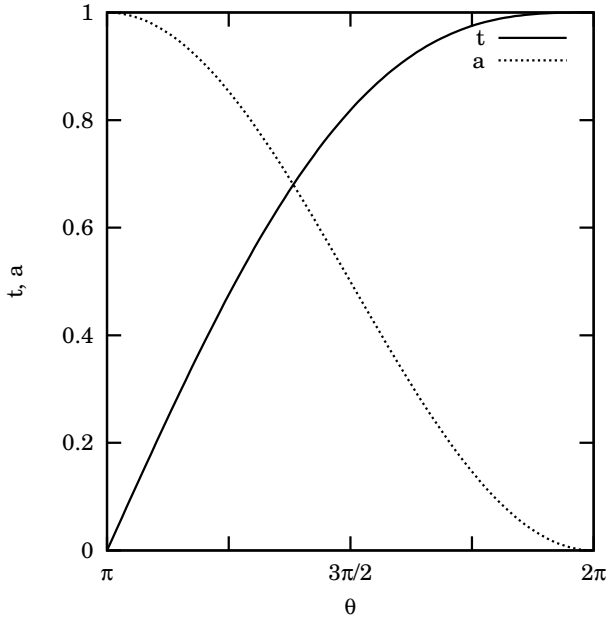}
 \caption{Normalized time $t$ and scale factor $a$ as a function of
   $\theta$ for the collapsing part ($\theta\in[\pi:2\pi]$) of the
   closed matter-dominated universe.}
\label{fig:time_theta}
\end{center}
\end{figure}

To find the solution, a parameterization is usually
proposed in which both the scale factor $a$ and time $t$ are functions
of a parameter $\theta$.  A usual parameterization is 
\begin{equation}
t \propto (\theta - \sin \theta)
\label{eq:t-theta}
\end{equation}
and
\begin{equation}
a(t) \propto (1- \cos \theta), 
\label{eq:a-theta}
\end{equation}
with $\theta\in[0:2\pi]$ \citep
{Narlikar1993}. In Figure \ref{fig:time_theta} we have plotted the time
and scale factor as a function of $\theta$ in normalized units for the
collapsing part of the evolution in this model---that is, for $\theta
\in[\pi:2\pi]$. This case has been studied by \citet{GP75},
who found the solution for $\delta$ as a function of $\theta$
as \citep[see also] [] {Narlikar1993}
\begin{equation}
\delta(\theta) = A \left[ \frac{5+\cos\theta}{1- \cos\theta} -
 \frac{3\theta \sin\theta}{(1-\cos\theta)^{2}}\right] + B
\frac{\sin\theta}{(1-\cos\theta)^{2}},
\label{eq:delta1}
\end{equation}
where $A$ and $B$ are constants, which can be evaluated using the
initial conditions as follows. We note that

%\noindent The collapsing part of this model corresponding to
%$\theta \in[\pi:2\pi]$. 
%
%
%$\delta(\theta=\pi)$ and
%$\dot{\delta}(\theta=\pi)$, to obtain a expressions for $A$ and $B$ as
%follows:
%
\begin{equation}
\delta_{\theta=\pi} = 2A
\label{eq:A}
\end{equation}
and
\begin{equation}
\left(\frac{d\delta}{d\theta}\right)_{\theta=\pi} =
\frac{3\pi A-B}{4}. 
\label{eq:AB}
\end{equation}

For $B > 0$, the last term in eq.\ (\ref{eq:delta1}) is positive but
monotonically decreasing in the interval $0 < \theta < \pi$,
approaching $+ \infty$ as $\theta \rightarrow 0+$, and becoming zero
at $\theta = \pi$. For $B < 0$, the sings are reversed, but the
divergence of this term as $\theta \rightarrow 0$ persists. For $\pi <
\theta < 2 \pi$, it approaches $\infty$ as $\theta \rightarrow 2
\pi$. On the other hand, the first term in the right-hand side of eq.\
(\ref{eq:delta1}) is zero at $\theta=0$. This implies that the full
solution with $B \ne 0$ diverges there. For the cosmological case,
which starts contracting at $\theta = \pi$, but must have previously
undergone an expanding stage ($0 < \theta < \pi$), we must require
that $\delta \rightarrow 0$ as $\theta \rightarrow 0+$ (the Big Bang),
and thus we must take $B=0$. The constant $A$ is determined by eq.\
(\ref{eq:A}), which gives $A=\delta_{\theta=\pi}/2$. Defining
$\delta_{\theta=\pi} \equiv \delta_{0} \equiv \rho_1(t=0)/\rho_0$, we
can then write the evolution equation for the cosmological case as
\begin{equation}
\deltac(\theta) = \frac{\delta_{0}}{2} \left[ \frac{5+\cos\theta}{1-
   \cos\theta} - \frac{3\theta \sin\theta}{(1-\cos\theta)^{2}}\right],
% + \frac{3\pi \delta_{0}}{2}\frac{\sin\theta}{(1-\cos\theta)^{2}}.
\label{eq:delta_ihf_cos}
\end{equation}
together with the initial condition, from eq.\ (\ref{eq:AB}),
\begin{equation}
\left. \dot \deltac \right|_{\theta=\pi} = \frac{3\pi \delta_0}{8},
\label{eq:ABbis}
\end{equation}
where we have defined $\dot \delta \equiv d\delta/d\theta$.

On the other hand, for the molecular cloud case, for which there is no
constraint that our initial state ($\theta = \pi$) be the continuation
of a previous expanding stage, we have no boundary condition applicable
at $\theta =0$, and thus no reason to set $B=0$. In this case, a
reasonable limiting initial condition is to set $\dot \delta = 0$ at
$\theta = \pi$, meaning that the fluctuation starts growing from
rest. From eq.\ (\ref{eq:AB}), this implies $B=3 \pi \delta_{0}/2$, and
thus the evolution equation for the perturbation becomes
\begin{equation}
\deltamc(\theta) = \frac{\delta_{0}}{2} \left[ \frac{5+\cos\theta}{1-
   \cos\theta} - \frac{3\theta \sin\theta}{(1-\cos\theta)^{2}}\right]
+ \frac{3\pi \delta_{0}}{2}\frac{\sin\theta}{(1-\cos\theta)^{2}},
\label{eq:delta_ihf_MC}
\end{equation}
with the initial condition
\begin{equation}
\left. \dot \deltamc \right|_{\theta=\pi} = 0.
\label{eq:ABtris}
\end{equation}
The solution starting from this initial condition should be considered
as describing the minimum possible growth rate of the density
fluctuation. In reality, the buildup of the fluctuation will imply that
at the initial time its growth rate is moderate but larger than
zero. However, since there is no criterion to decide the initial growth
rate, we consider the case of zero rate as the lower limit to the
possible realistic rates, and so the evolution of actual fluctuations
should be considered to be bounded from below by this case.

Equations (\ref{eq:delta_ihf_cos}) and (\ref{eq:delta_ihf_MC}), with the
corresponding initial conditions given by eqs.\ (\ref{eq:ABbis}) and
(\ref{eq:ABtris}), describe the evolution of a density
fluctuation in a contracting background, in the cosmological and
molecular cloud subcases, respectively. In general, we will refer to the
situation of a contracting background, as an Inverse Hubble Flow (IHF).

Finally, we note that the collapsing background (assumed spherical) for
either IHF case completes its collapse on its free-fall time, which is
given by

\begin{equation}
\tauff = \sqrt{\frac{3 \pi}{32 G \rho_{0}}}.
\label{eq:tff}
\end{equation}
Because at $t = \tauff$ the scale factor of the background $a$ has
shrunk to zero, $\tauff$ corresponds to $\theta=2 \pi$. On the
other hand, we consider that the evolution starts when $\theta =
\pi$, that is, at the onset of the contracting stage. Thus,
relation (\ref{eq:t-theta}) can be written as the equality 
\begin{equation}
t = \left(\frac{\theta - \sin \theta} {\pi} - 1\right)\, \tauff.
\label{eq:t-theta_full}
\end{equation}

\subsection{The Jeans case: a static background} \label{sec:Jeans}

In this next section, we will compare the evolution of the
fluctuations in an IHF to the evolution of the classical Jeans case,
applicable to a static background. For this case, it is well
known\citep[see, e.g.,] [] {Binney1987} that the general solution is
\begin{equation}
\delta_{\mathrm{J}} (t) = \alpha e^{t/\tau} + \beta e^{-t/\tau},
\label{eq:delta_jeans}
\end{equation}
where the characteristic timescale $\tau$ is given by
\begin{equation}
\tau = \sqrt{\frac{1}{4 \pi G \rho_{0}}} = \sqrt{\frac {8} {3 \pi^2}}
\, \tauff,
\end{equation}
and $\alpha$ and $\beta$ are coefficients to be determined as follows.
First, we note that, to compare to the cosmological IHF case, the
simplest choice is $\alpha = \delta_0$ and $\beta = 0$, since we then
have 
\begin{equation}
\left(\frac{d \delta_{\rm J}} {dt} \right)_{t=0} = \frac{\delta_0} {\tau}.
%\left.\dot \delta_{\rm J} \right|_{t=0} =  \frac{\delta_0} {\tau}.
\end{equation}
This can be compared to the initial growth rate for the cosmological IHF
case which, writing eq.\ (\ref{eq:ABbis}) in terms of the time
variable, is
\begin{equation}
\left(\frac{d \delta_{\rm c}} {dt} \right)_{t=0}  =  \frac{3 \pi} {8}
\left(\frac{2}{3} \right)^{1/2} \frac{\delta_0} {\tau} \approx 0.962
\, \frac{\delta_0} {\tau},
%\left.\dot \deltac \right|_{t=0} =  \frac{3 \pi} {8}
%\left(\frac{2}{3} \right)^{1/2} \frac{\delta_0} {\tau} \approx 0.962
%\, \frac{\delta_0} {\tau},
\end{equation}
and so the two initial growth rates are nearly the same.

On the other hand, in order to compare to the molecular-cloud IHF
solution, a useful choice is $\alpha = \beta = \delta_0/2$, since
in this case the initial growth rate for the Jeans solution is zero, in
agreement with the initial condition for the molecular-cloud IHF
solution.

\section{Results} \label{sec:results}

\begin{figure}%[p]
\begin{center}
  \includegraphics[width=1.0\linewidth]{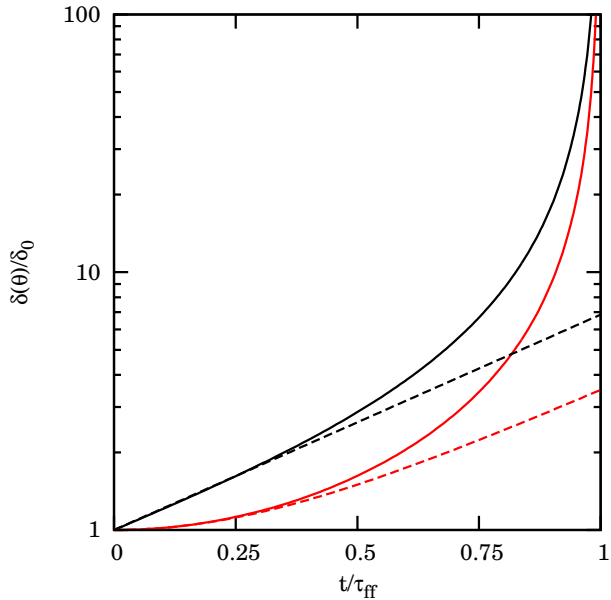} 
  \caption{Linear growth of the perturbations ($\delta/\delta_{0}$) in
    an inverse Hubble flow (IHF; solid lines) and in a static
    background (the Jeans case; dashed lines) as a function of time,
    as given by eq.\ (\ref{eq:t-theta_full}). The cosmological case,
    given by eqs.\ (\ref{eq:delta_ihf_cos}) and (\ref{eq:ABbis}) for
    the IHF flow, and by eq.\ (\ref{eq:delta_jeans}) with
    $\alpha=\delta_0$ and $\beta=0$ for the Jeans case, is shown by
    the black lines. The molecular cloud case, given by eqs.\
    (\ref{eq:delta_ihf_MC}) and (\ref{eq:ABtris}) for the IHF, and by
    eq.\ (\ref{eq:delta_jeans}) with $\alpha = \beta = \delta_0/2$ for
    the Jeans case, is shown by the red lines. The timescale in all
    cases is given in units of the free-fall time for the IHF
    background, eq.\ (\ref{eq:tff}).}
\label{fig:deltas}
\end{center}
\end{figure}

Equations (\ref{eq:delta_ihf_cos}) and (\ref{eq:delta_ihf_MC}) on one
hand, and eq.\ (\ref{eq:delta_jeans}) on the other, describe the
linear growth of the density perturbations in
%a collapsing spherical MC
an IHF and in the Jeans' case, respectively. Note that these expressions
contain an explicit dependence on the initial density perturbation
$\delta_{0}$.

%In the IHF case, the linear perturbation inside the MC is
%accreting material from the parent cloud (a collapse-within-collapse,
%CWC). 

In Figure \ref{fig:deltas}, we plot the two IHF solutions, eqs.\
(\ref{eq:delta_ihf_cos}) and (\ref{eq:delta_ihf_MC}) (black and red
solid lines, respectively) as a function of time, together with the
Jeans solution with either finite ($\alpha=\delta_{0}$,
  $\beta=0$) or zero ($\alpha=\beta=\delta_{0}/2$) initial growth
rate (black and red dashed lines, respectively). The
  timescale for all cases is normalized to the free-fall time,
$\tauff$, and all solutions are normalized to the initial
fluctuation amplitude, $\delta_{0}$.
%For comparison, we also
%plot in Figure~\ref{fig:deltas} the density growth for a collapsing
%self-graviting spherical cloud (dash-dotted line), where $\rho_{0}$ is
%the initial volume density. Note that we have to plot the absolute
%density $\rho$, rather than the density fluctuation $\delta$, since
%there is no background density to refer to. However, the density is
%still normalized to its initial value, $\rho_0$. 
%
From Figure \ref{fig:deltas}, the linear growth of the normalized
density fluctuation is seen to be 
%much 
faster in the IHF case{\bf s} than in
the Jeans case. Moreover, the fact that $\delta$ increases
monotonically with time means that the density of the perturbation
increases faster than that of the collapsing background.

However, note that, for the linear IHF solutions, the fluctuation
terminates its collapse at the same time as the background, since,
from eqs.\ (\ref{eq:delta_ihf_cos}) and (\ref{eq:delta_ihf_MC}), it is
seen that $\delta \rightarrow \infty$ as $\theta \rightarrow 2 \pi$,
as seen in Fig.\ \ref{fig:deltas}. The anticipated collapse of
the fluctuation with respect to that of the background occurs as a
consequence of the {\it nonlinear} growth of the fluctuation, which
starts when $\delta \approx 1$. After this time, the fluctuation
collapses on its own free-fall time and, since its density is now
roughly twice that of the background at this time, its free-fall time
is now $\sim 1/\sqrt{2}$ that of the background, also at this time,
implying an anticipated collapse of the fluctuation.

\begin{figure}%[p]
\begin{center}
 \includegraphics[width=1.0\linewidth]{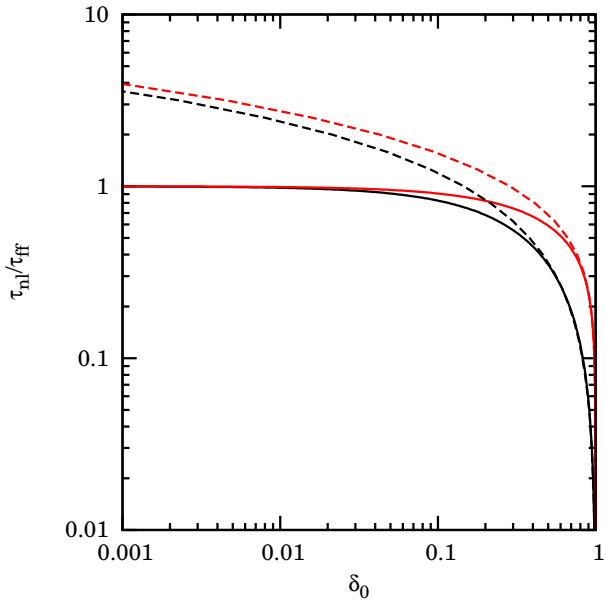}
 \caption{Time $\taunl$ for the perturbation to grow to
     $\delta=1$ in both the Jeans (dashed lines) and the IHF (solid
   lines) cases, {\it versus} the initial fluctuation amplitude,
   $\delta_0$. As in Fig.\ \ref{fig:deltas}, the black lines
     denote the cosmological case, and the red lines denote the
     molecular cloud case.}
\label{fig:times}
\end{center}
\end{figure}

Figure \ref{fig:times} illustrates the time for the perturbation to
become nonlinear ($\taunl$), that is, the time for the perturbation to
grow to $\delta=1$, as a function of the initial fluctuation
amplitude, $\delta_{0}$. For example, consider the case of $\delta_{0}
= 10^{-1}$. The time for the density fluctuation to grow by a factor
of $1/\delta_{0}$ (i.e., the time to reach nonlinearity, or $\delta =
1$) in the cosmological IHF case and the corresponding Jeans case is
$\approx 0.8$ and $1.2 \tauff$, respectively, as shown by the black
solid and dashed lines, respectively.  For the molecular cloud case
and its correponding Jeans case with zero initial growth rate (red
solid and dashed lines, respectively), these times are $\sim 0.9$ and
$\sim 1.5 \tauff$, respectively.  In general, this figure shows that,
the smaller the initial amplitude, the larger the ratio of the Jeans
to the IHF growth times. More fundamentally, the linear growth time in
the Jeans case increases without limit as $\delta_0$ decreases, while
it remains finite, asymptotically approaching the free-fall time for
the IHF cases. Again, this is a manifestation of the inherent
nonlinearity of the situation when a small-scale perturbation is
growing within a larger-scale one that is also growing.

\section{Discussion} \label{sec:disc}

The results from the previous section have a number of important
implications for the evolution of density fluctuations within
collapsing MCs. First and foremost, our results imply that the
standard notion that, in the linear regime, all fluctuations grow at
the same rate, is not entirely accurate.  Small-scale fluctuations
embedded in a larger-scale fluctuation that is also collapsing will
grow faster than isolated fluctuations. The fact that this feature has
not been recognized before can be traced to the common practice of
discussing purely in terms of Fourier spectra, neglecting the {\it
  phases} of the Fourier modes, which are the part related to the
spatial distribution of the density field \citep{AF85}.  In {\bf our}
case, it is seen that the spatial location introduces a fundamental
modification to the picture. This coupling between a small and a large
scale is inevitable even in the linear regime, and arises from the
fact that the large-scale fluctuation embedding the small-scale one
constitutes a background which modifies the governing equations,
even in the linear analysis.

In the case of structure formation within the expanding cosmological
flow, it is well known that all density fluctuations are initially part
of the expanding Universe, and therefore, although their corresponding
density fluctuation $\delta$ is increasing monotonically, their physical
size is still increasing as a consequence of the expansion, albeit at a lower
rate than the global one. At some time, usually referred to as the {\it
turnaround point}, they finally stop expanding and begin to contract,
but by this time they are already nonlinear with respect to the
universal Hubble flow. Therefore, it is possible that the {\it linear} growth
of a small-scale fluctuation embedded within a larger-scale one will
occur while the latter is still expanding, even if its corresponding
$\delta$ is increasing. In this case, the small-scale fluctuation will
nevertheless find itself embedded in a background that is expanding more
slowly that the global expansion, and so its growth will be faster than
that of a fluctuation growing in the global expansion field, causing an
earlier collapse anyway. We hope to
investigate this in future work.

In fact, it is possible that the accelerated growth of density
fluctuations embedded in growing density fluctuations may have already
been observed in numerical simulations of halo and galaxy formation,
but not recognized as a consequence of this effect. Indeed, the
numerical study of the environmental dependence of halo formation by
\citet{ST2004} found evidence that {\it halos of a given mass forming
  in high density environments typically form earlier than those
  formed in low density environments}. This result has been
subsequently confirmed by other authors \citep[e.g.,][]{Harker2006,
  Wechsler2006, Maulbetsch2007, Wang2011}, some of which use a
different measure of the environment's overdensity.\footnote{Note,
  however, that this effect is weak compared to the dependence on the
  mass assembly history.}  We suggest that this result may be a
consequence of the halo being located in an overdense region which is
itself growing, therefore, it grows faster than otherwise.

In the MC context, our results have also relevant implications. First,
if the formation mechanism of MCs is such that they quickly acquire
many Jeans masses, then their collapse is nearly pressureless
\citep[as suggested by many numerical studies; see, e.g.,] [] {VS+07,
  VS+11, HH08, Heitsch+08, Heitsch+09, Banerjee+09, Micic+13}, and
then even moderately linear perturbations, with $\delta_0 \gtrsim
10^{-1}$ can grow faster than the global collapse, as systematically
observed in those simulations. Of course, this is supplemented by the
facts that a) the turbulence may produce moderately {\it nonlinear}
fluctuations, which then have shorter free-fall times even from the
start, and b) that flattened or filamentary structures have longer
free-fall times than spherical ones, so that spheroidal fluctuations
within such structures will collapse earlier than the larger structure
\citep{Toala+12, Pon+12}.  But the important issue here is that, even
in the worst-case scenario for fragmentation, namely that of a
spherical geometry with linear density fluctuations, the perturbations
are able to collapse earlier than the cloud.

\section{Summary and Conclusions} \label{sec:concl}

Recent numerical simulations of MC formation and evolution have shown
that the clouds engage in global gravitational collapse, and that the
density fluctuations within them grow and complete their local
collapse before the global collapse is completed \citep{VS+07, VS+11,
  HH08, Heitsch+08, Heitsch+09, Banerjee+09, Micic+13}. Motivated by
these results, in this paper we have performed a linear analysis of
the growth rate of density fluctuations immersed in a medium that is
itself undergoing global gravitational contraction. We have used the
standard linear analysis used for the growth of density fluctuations
in Hubble flows, but considering the case of an {\it inverse} Hubble
flow (IHF), where the background is contracting rather than
expanding. We considered two variants of an IHF. The first is
appropriate for a cosmological setup, in which the initial fluctuation
at the onset of the collapse of the background is already growing at a
finite rate, consistent with the evolution during a previous epoch of
growth, during the epoch when the background was still expanding. The
other is meaningful as a lower limit for a collapsing molecular cloud,
in which we assume the initial growth rate to be zero. This represents
a lower limit to the possible initial growth rate of the fluctuation,
which is unconstrained in this case.

Our main results are the following:

\begin{itemize}

\item Density fluctuations embedded in an IHF grow faster than in the
standard Jeans analysis (where the background is static). 
%In fact, their
%growth rate is even larger than the free-fall rate for a uniform sphere.

\item While in the Jeans case the growth time to reach nonlinearity
(i.e., to reach a density fluctuation amplitude $\delta =1$) increases
without limit as $\delta_0 \rightarrow 0$, where $\delta_0$ is the
initial value of $\delta$, in the IHF case this growth time
is bounded from above by the free-fall time of the background density,
$\tauff$. This reflects the fact that the density fluctuation is
embedded in a medium which is itself collapsing on a timescale
$\tauff$, and the longest possible time for the fluctuation to complete
its collapse is the free-fall time of the medium in which it is
embedded.

\item For perturbations embedded in an IHF having initial amplitudes
$\delta_0 \gtrsim 10^{-1}$, the time to reach nonlinearity is $\sim 0.8
\tauff$ in the cosmological variant, and $\sim 0.9 \tauff$ in the
molecular cloud variant. From then on, the perturbation grows at its own
free-fall rate, and therefore it will always be ``ahead'' of the global
collapse.

\end{itemize}

In the context of large-scale structure formation in the Universe, these
results may offer an alternative (or complementary) explanation for the
observation in numerical simulations that halos of a given mass
typically form earlier in high density environments than those formed in
low density environments. In the MC context, they may offer an
explanation for the ubiquitous observation in numerical simulations that
small-scale density fluctuations within large MCs complete their
collapse (i.e., form stars) before the collapse of the parent cloud is
completed, even if the time difference is not too large. It must
also be borne in mind that the spherical symmetry that we have assumed
for the background constitutes a worst-case scenario for the possibility
of fragmentation, since in this case the free-fall time depends only on
the density, and is independent of size scale. Instead, for flattened or
filamentary geometries, the actual collapse time is larger than the
spherical free-fall time by factors that depend on the aspect ratio of
the object \citep{Toala+12, Pon+12}. 

More fundamentally, our results show that the hierarchical nesting
of the fluctuations of different scales affects the growth rate of
density fluctuations, even in the linear regime. Thus, our results imply
that density perturbations are always able to collapse earlier than the
whole cloud, as envisioned by \citet{Hoyle1953}, and recently observed
in the numerical simulations, regardless of whether they have linear or
nonlinear amplitudes, making Hoyle-like fragmentation an inescapable
process in multi-Jeans-mass molecular clouds.

\section*{Acknowledgments}

We would like to thank the Scientific Editor for helpful comments.
J.A.T. acknowledges support by the CSIC JAE-Pre student grant
2011-00189. E.V.-S.\ acknowledges financial support from CONACYT grant
102488. G.C.G. acknowledges financial support from PAPIIT grant
IN111313.

\appendix

\section[]{Friedmann's Equation for a Spherical Molecular Cloud}

Let us suppose a spherical cloud with total mass $M$ and initial
radius $R(t=0)=R_{0}$. By energy conservation we have
\begin{equation}
\frac{1}{2}v(t)^{2} - \frac{GM}{R(t)} = E_{\mathrm{TOT}},
\label{eq:energia}
\end{equation}
\noindent and $E_{\mathrm{TOT}}$ can be evaluated at $t=0$, supposing
that $v(t)=0$. This gives
\begin{equation}
E_{\mathrm{TOT}} = - \frac{GM}{R_{0}}.
\end{equation}
\noindent Manipulating Ec.~(\ref{eq:energia}) we can write
\begin{equation}
\frac{v(t)^{2}}{R(t)^{2}} - \frac{2GM}{R(t)^{3}} = -
\frac{2GM}{R_{0}}\frac{1}{R(t)^{2}},
\end{equation}
\noindent and using $M/R(t)^{3}=4/3 \pi \rho(t)$, we can rewrite the
former equation as
\begin{equation}
\frac{v(t)^{2}}{R(t)^{2}} - \frac{8 \pi G \rho(t)}{3} = -\frac{8 \pi G
 \rho_{0}}{3}\frac{R_{0}^{2}}{R(t)^{2}}.
\label{eq:energia1}
\end{equation}

If we now define $a(t) = R(t)/R_{0}$, this is,
$\dot{a(t)}=\dot{R}(t)/R_{0}$ with $v(t)=\dot{R}(t)$, we can write
Eq.~(\ref{eq:energia1}) as
\begin{equation}
\left(\frac{\dot{a}}{a}\right)^{2} + \frac{8 \pi G
 \rho_{0}}{3}\frac{1}{a^2} = \frac{8 \pi G \rho(t)}{3}.
\end{equation}
Under this scenario, we can define $k\equiv 8/3 \pi G \rho_{0}$ and
write
\begin{equation}
\left(\frac{\dot{a}}{a}\right)^{2} + \frac{k}{a^2} = \frac{8 \pi G
 \rho(t)}{3}.
\end{equation}

\end{document}